\documentclass[aps,twocolumn,floats,prd,nofootinbib,10pt,longbibliography,superscriptaddress]{revtex4-1}

\usepackage[dvips]{graphicx} % 
\usepackage{graphicx,amsmath,amsfonts,amssymb,slashed,float,hyperref}
\usepackage[normalem]{ulem}
\usepackage{bbold,wasysym}
\usepackage{graphicx}
\usepackage{array,multirow}
\usepackage[utf8]{inputenc}

\usepackage[usenames,dvipsnames]{xcolor} 

\usepackage{soul}

\definecolor{RedWine}{rgb}{0.743,0,0}
\definecolor{RoyalBlue}{rgb}{0.25,.41,.88}
\definecolor{celestialblue}{rgb}{0.29, 0.59, 0.82}

\setstcolor{Blue}

\newcommand{\LCDM}{$\Lambda$CDM}

\begin{document}

\title{Early dark energy is not excluded by current large-scale structure data}

\author{Tristan L.~Smith}
\affiliation{Department of Physics and Astronomy, Swarthmore College, Swarthmore, PA 19081, USA}
\author{Vivian Poulin}
\affiliation{Laboratoire Univers \& Particules de Montpellier (LUPM),
CNRS \& Universit\'{e} de Montpellier (UMR-5299),
Place Eug\`{e}ne Bataillon, F-34095 Montpellier Cedex 05, France}
\author{Jos\'e Luis Bernal}
\affiliation{Department of Physics and Astronomy, Johns Hopkins University, Baltimore, MD 21218, USA}
\author{Kimberly K.~Boddy}
\affiliation{Theory Group, Department of Physics, The University of Texas at Austin, Austin, TX 78712, USA}
\author{Marc Kamionkowski}
\affiliation{Department of Physics and Astronomy, Johns Hopkins University, Baltimore, MD 21218, USA}
\author{Riccardo Murgia}
\affiliation{Laboratoire Univers \& Particules de Montpellier (LUPM),
CNRS \& Universit\'{e} de Montpellier (UMR-5299),
Place Eug\`{e}ne Bataillon, F-34095 Montpellier Cedex 05, France}

\begin{abstract}
We revisit the impact of early dark energy (EDE) on galaxy clustering using BOSS galaxy power spectra, analyzed using the effective field theory (EFT) of large-scale structure (LSS), and anisotropies of the cosmic microwave background (CMB) from {\it Planck}.
Recent studies found that these data place stringent constraints on the maximum abundance of EDE allowed in the Universe.
We argue here that their conclusions are a consequence of their choice of priors on the EDE parameter space, rather than any disagreement between the data and the model. For example, when considering EFT-LSS, CMB, and high-redshift supernovae data we find the EDE and \LCDM{} models can provide statistically indistinguishable fits ($\Delta \chi^2  = 0.12$) with a relatively large value for the maximum fraction of energy density in the EDE ($f_{\rm ede} = 0.09$) and Hubble constant ($H_0 = 71$ km/s/Mpc) in the EDE model. 
Moreover, we demonstrate that the constraining power added from the inclusion of EFT-LSS traces to the potential tension between the power-spectrum amplitudes $A_s$ derived from BOSS and from \textit{Planck} that arises even within the context of $\Lambda$CDM.  Until this is better understood, caution should be used when interpreting EFT-BOSS+{\it Planck} constraints to models beyond \LCDM{}. 
These findings suggest that EDE still provides a potential resolution to the Hubble tension and that it is worthwhile to test the predictions of EDE with future data-sets and further study its theoretical possibilities.  
\end{abstract}
\date{\today}

\maketitle

\section{Introduction}
\label{sec:into}

Over the past several years, the standard cosmological model \LCDM{} has come under increased scrutiny as measurements of the late-time expansion history of the Universe~\cite{Scolnic:2017caz}, the cosmic microwave background (CMB)~\cite{Aghanim:2018eyx}, and large-scale structure (LSS)---such as the clustering of galaxies~\cite{Alam:2016hwk,Abbott:2017wau,Hildebrandt:2018yau,eBOSS_cosmo}---have improved. 
Observations have spurred recent tensions within \LCDM{}, related to the Hubble constant $H_0 = 100 h$ km/s/Mpc~\cite{Verde:2019ivm} and the parameter combination $S_8 \equiv \sigma_8(\Omega_{\rm m}/0.3)^{0.5}$~\cite{Joudaki:2019pmv} (where $\Omega_{\rm m}$ is the total matter relic density and $\sigma_8$ is the variance of matter perturbations within 8 Mpc/$h$ today), reaching the $\sim 5\sigma$ and $3\sigma$ level, respectively.

Modifications of the low redshift Universe are unlikely to provide a satisfactory resolution~\cite{Bernal:2016gxb,DiValentino:2017zyq,Addison:2017fdm,DiValentino:2017iww,Verde:2016ccp,Poulin:2018zxs,Aylor:2018drw,Arendse:2019hev,2020PhRvD.101h3524R} to the $H_0$ tension. Thus, efforts have shifted towards pre-recombination modifications of \LCDM{}~\cite{Knox:2019rjx}.
Models of early dark energy (EDE), in particular, have shown promise (e.g., \cite{Karwal:2016vyq,Poulin:2018cxd,Lin:2019qug,Smith:2019ihp,Berghaus:2019cls,Niedermann:2019olb,Sakstein:2019fmf,Niedermann:2020dwg}).

The EDE resolution to the $H_0$ tension also makes unique predictions for the observed LSS. 
Recent work has explored the impact of EDE on weak-lensing observations~\cite{Hill:2020osr} and galaxy clustering~\cite{Ivanov:2020ril,DAmico:2020ods}, reporting increasingly tight constraints on the maximum fraction of the total energy density of the Universe in EDE, $f_{\rm ede}$: $f_{\rm ede} < 0.05$ at the 95\% confidence level (CL)
for an EDE model with three free parameters.\footnote{EDE models are generally specified by four parameters: $f_{\rm ede}$, the redshift $z_c$ at which the maximum of the EDE contribution to the total energy density occurs, the initial field displacement, $\Theta_i$, and the potential's power-law index around its minimum, $n_{\rm axion}$, fixed to $n_{\rm axion}=6$ in Refs.~\cite{Hill:2020osr,Ivanov:2020ril,DAmico:2020ods} and in this work. See Ref.~\cite{Smith:2019ihp} for more details.} With such a small upper limit, these papers claim to have effectively ruled out the EDE scenario as a resolution to the Hubble tension.

In this paper, we reconsider the constraints on EDE from BOSS galaxy clustering observations, analyzed using the effective field theory (EFT) of large-scale structure \cite{Baumann:2010tm, Carrasco:2012cv,Colas:2019ret,DAmico:2019fhj,Ivanov:2019pdj}.
First, using a three-parameter EDE model, we confirm the results from previous studies \cite{Ivanov:2020ril,DAmico:2020ods}, and find that the inclusion of the EFT-BOSS data constrains $f_{\rm ede} < 0.053$ at the 95\% CL.\footnote{The upper limit we find is slightly smaller than the one reported in Ref.~\cite{DAmico:2020ods}. This is most likely due to the fact that our chains have a more stringent convergence requirement ($R-1<0.03$).}  However, we disagree with their conclusions that this upper limit effectively rules out EDE as a resolution to the Hubble tension. We make a distinction between the posterior distribution for $f_{\rm ede}$ given a choice of priors and the fact that there are parameter values with $f_{\rm ede}$ much greater than this limit which provide a fit to the data that is statistically indistinguishable from \LCDM{}. In order to demonstrate this, using the same approach in Ref.~\cite{vivian}, we consider a one-parameter EDE model (1pEDE) in which the only free EDE parameter is $f_{\rm ede}$. In this case we find that $f_{\rm ede} = 0.0523^{+0.026}_{-0.036}$, with a 95\% CL upper limit of $f_{\rm ede} < 0.107$. This choice of EDE prior provides a proof of principle that there are EDE parameter values which fit the data well and are much larger than the previously reported upper limits. Further exploring this, we find that an EDE model with $f_{\rm ede} = 0.09$ and $h=0.71$ can fit these data as well as \LCDM.

We confirm that the inclusion of the EFT-BOSS data leads to a tighter constraint on $f_{\rm ede}$, even when considering the 1pEDE. Further exploration of where this additional constraining power comes from points to the potential inconsistency between the value of the scalar amplitude, $A_s$, inferred from the EFT-BOSS data and from {\it Planck} data. The positive correlation between $f_{\rm ede}$ and $A_s$ then tends to decrease the allowed values of $f_{\rm ede}$. Since the mismatch in $A_s$ also occurs in \LCDM{}, we argue that one should be cautious when interpreting constraints to models beyond \LCDM{} obtained by combining {\it Planck} and EFT-LSS data. 

Using a similar approach as presented here, a reassessment of weak-lensing observations in the context of EDE \cite{vivian} also found that these constraints are not robust to the choice of EDE priors, and that when the parameter space is reduced to just $f_{\rm ede}$, the constraints relax to $f_{\rm ede}<0.094$ at 95\% CL. Moreover, the apparent constraining power is entirely driven by a $\sim3\sigma$ statistical inconsistency that is already present between joint KiDS+Viking+DES data \cite{Asgari:2019fkq} and the $\Lambda$CDM model inferred from \textit{Planck} data, which makes it hard to properly interpret constraints to beyond-\LCDM{} models when using these data. 

This paper is organized as follows: we present the analysis method in Sec.~\ref{sec:analymeth}; discuss the consequences of the parametrization of EDE models on the final parameter inference and propose using only one EDE free parameter in Sec.~\ref{sec:EDEparam}; explore the additional constraining power that the galaxy power spectrum measurements provide to EDE analyses in Sec.~\ref{sec:EDELSS}; and conclude in Sec.~\ref{sec:cons}.

\section{Analysis method}
\label{sec:analymeth}

We run a Markov-chain Monte Carlo (MCMC) using the public code {\sf MontePython-v3}\footnote{\url{https://github.com/brinckmann/montepython_public}} \citep{Audren:2012wb,Brinckmann:2018cvx}, interfaced with our modified version of {\sf CLASS}\footnote{\url{https://github.com/PoulinV/AxiCLASS}}.
We perform the analysis with a Metropolis-Hasting algorithm, assuming flat priors on $\{\omega_b,\omega_{\rm cdm},\theta_s,A_s,n_s,\tau_{\rm reio}\}$; when considering the three parameter EDE (3pEDE) model we also vary $\{\log_{10}(z_c),f_{\rm ede} ,\Theta_i\}$ and for the one parameter model (1pEDE) we fix $\log_{10}(z_c) = 3.569$ and $\Theta_i = 2.775$ (their best-fit values for the 3pEDE using only {\it Planck} power spectra \cite{vivian}). As described in Ref.~\cite{Smith:2019ihp}, we use a shooting method to map the set of phenomenological parameters $\{\log_{10}(z_c), f_{\rm ede}\}$ to the theory parameters $\{m,f\}$. 
We adopt the {\em Planck} collaboration convention and model free-streaming neutrinos as two massless species and one massive with $m_\nu=0.06$ eV \cite{Ade:2018sbj}. 
We consider chains to be converged using the Gelman-Rubin \citep{Gelman:1992zz} criterion $R -1<0.03$. To post-process the chains and produce our figures we used {\sf GetDist} \cite{Lewis:2019xzd}.

\begin{figure}
\centering
\includegraphics[width=\columnwidth]{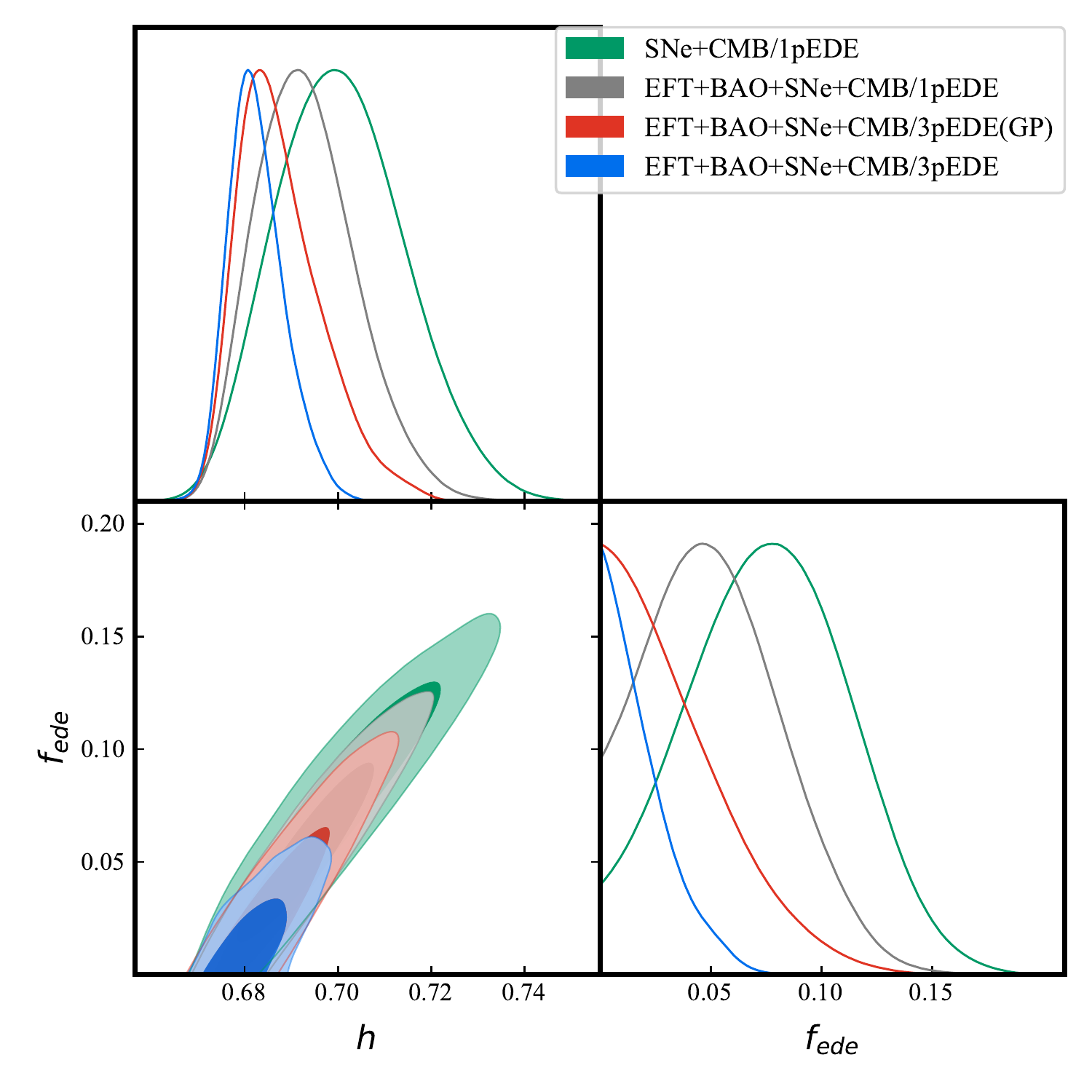}
\caption{68\% and 95\% CL marginalized constraints on the $h$-$f_{\rm ede}$ plane for different EDE parametrizations and data combinations, as described in the legend. `(GP)' denotes an analysis in which we place Gaussian priors on $\log_{10}(z_c) = 3.5\pm 0.1$ and $\Theta_i <3.1$ with $\sigma_\Theta = 0.5$.}
\label{fig:progressive_freedom}
\end{figure}

We use the EFT `full shape' analysis of Refs.~\cite{DAmico:2019fhj,Colas:2019ret,DAmico:2020kxu} applied to the pre-reconstructed BOSS galaxy clustering measurements presented in Refs.~\cite{Gil-Marin:2015sqa,Gil-Marin:2015nqa, DAmico:2020kxu}\footnote{\url{https://pybird.readthedocs.io/en/latest/}}, as well as the post-reconstructed anisotropic BAO measurements of BOSS DR12 at $z=0.32$ and $0.57$ \cite{Alam:2016hwk} which correspond to the LOWZ and CMASS samples. Note that, as in Ref.~\cite{DAmico:2019fhj}, we do not use the south galactic cap field of LOWZ. We include a covariance between the EFT-BOSS and anisotropic BAO analysis \cite{DAmico:2020kxu}. In the following we will refer to the joint EFT-BOSS + BAO analysis as `EFT+BAO', the {\em Planck} CMB and lensing power spectra as `CMB', the Pantheon type Ia supernova dataset as `SNe', and the big bang nucleosynthesis (BBN) prior on $\omega_b$ \cite{Tanabashi:2018oca} as `BBN'.

\section{Impact of EDE prior}
\label{sec:EDEparam}

The EDE model considered here and in the LSS analyses in Refs.~\cite{Hill:2020osr,Ivanov:2020ril,DAmico:2020ods} is a phenomenological model that provides a physically consistent evolution of the background and perturbations of an EDE component which has a constant background energy density up until some critical redshift, $z_c$, and then becomes dynamical and dilutes, due to the expansion of the universe. After fixing the shape of the potential, this model is specified by the standard six \LCDM{} parameters plus three EDE-specific parameters (3pEDE), only one of which, $f_{\rm ede}$, controls the overall energy density in the EDE.

\begin{figure*}[t]
\centering
\includegraphics[width=\textwidth]{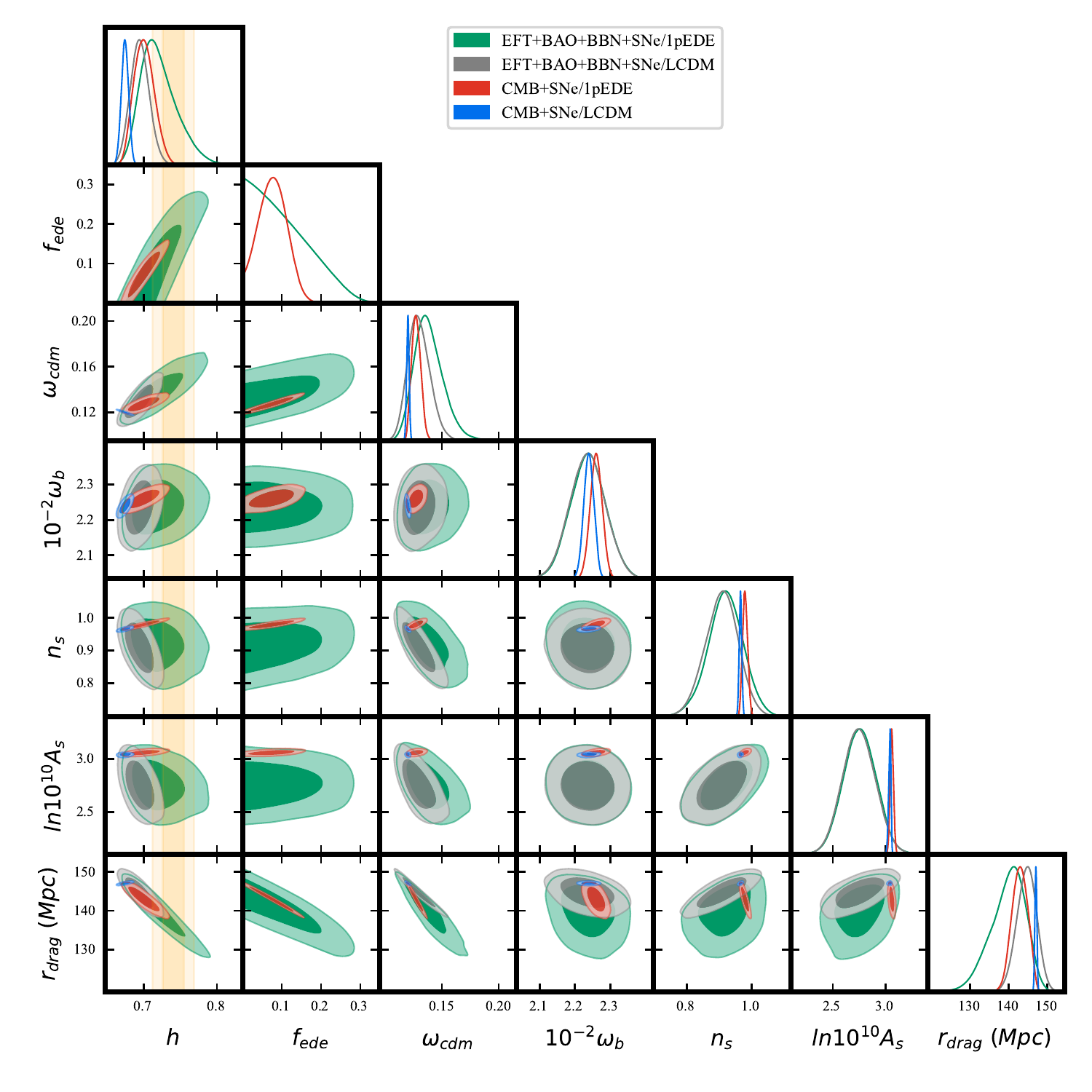}
\caption{ 68\% and 95\% CL marginalized constraints along with one-dimensional marginalized posteriors for parameters which correlate with $f_{\rm ede}$. We show results from EFT+BAO+BBN+SNe assuming 1pEDE (green) and \LCDM\ (gray), and similarly but using CMB+SNe (red and blue, respectively). The yellow band shows the 68\% (darker)/95\% (lighter) CL for $h$ determined by the SH0ES collaboration \cite{Riess:2019cxk}.}
\label{fig:separate}
\end{figure*}

As $f_{\rm ede}$ tends to zero, a change in the other EDE parameters has no measurable impact on the EDE predictions. This leads to a prior preference for this region of the parameter space (i.e., a large prior volume), as any point corresponds to similar likelihood values, hence the posterior density will be larger around these points. This issue arises in \emph{any} parameterization of the EDE model since the relation between the phenomenological and any other parameters (including the theoretical ones) is bijective, so that there will be some direction in this other parameter space which corresponds to $f_{\rm ede} \rightarrow 0$, again leading to the same prior preference.
Therefore, upper limits on $f_{\rm ede}$ assuming 3pEDE do not reflect that the same or higher likelihood can be achieved with a larger $f_{\rm ede}$. In previous work \cite{Poulin:2018cxd, Smith:2019ihp, Lin:2019qug, Lin:2020jcb} this issue was addressed by including the SH0ES prior, $H_0=74.03\pm1.42$ km/s/Mpc~\cite{Riess:2019cxk}, by placing a non-zero lower-limit, $f_{\rm ede} \geq 0.04$~\cite{Ivanov:2020ril}, or by imposing narrow Gaussian priors on $\log_{10}(z_c)$ and $\Theta_i$ \cite{DAmico:2020ods} (denoted by `(GP)' in Fig.~\ref{fig:progressive_freedom}). Instead, here we attempt to mitigate this by considering a one parameter EDE (1pEDE) model where we only allow $f_{\rm ede}$ to vary. 

First, when exploring that 3pEDE with the EFT+BAO+SNe+CMB datasets, we find posteriors on $f_{\rm ede}$ in agreement with Refs.~\cite{Ivanov:2020ril,DAmico:2020ods}: $f_{\rm ede} < 0.053$ at 95\% CL. However, using the same data for the 1pEDE leads to $f_{\rm ede} = 0.0523_{-0.036}^{+0.026}$ with a 95\% CL upper limit $f_{\rm ede} < 0.107$. Fig.~\ref{fig:progressive_freedom} indicates that by reducing the number of EDE parameters we have weakened the constraint on $f_{\rm ede}$. Although this effect may appear counterintuitive (in most cases, having more free parameters weakens parameter constraints), once we note that $f_{\rm ede}$ is not correlated with the rest of EDE parameters in the region of interest of the parameter space,
using 1pEDE gives a more direct exploration of the likelihood dependence on $f_{\rm ede}$. Moreover, it is clear from Fig.~\ref{fig:progressive_freedom} that using narrower priors on $\log_{10}(z_c)$ and $\Theta_i$, 
as in Ref.~\cite{DAmico:2020kxu}, is not sufficient to uncover the likelihood's dependence on $f_{\rm ede}$. 
Similar results are obtained with the implementation of EFT-BOSS from Ref.~\cite{Ivanov:2020ril}, as we find that in the 1pEDE model, the combination of EFT+BAO+CMB leads to $f_{\rm EDE} =  0.072\pm0.034$, with a 95\% CL upper limit $f_{\rm ede} < 0.132$. 

In order to demonstrate that the EDE model, with relatively large values of $f_{\rm ede}$ and $H_0$, can fit the EFT+BAO+SNe+CMB data as well as \LCDM{}, we searched for the minimum $\chi^2$, fixing $H_0 = 71$ km/s/Mpc for the 3pEDE model. The differences in the fit, $\chi^2_{\rm EDE} - \chi^2_{\Lambda {\rm CDM}} = 0.12$, make the models statistically indistinguishable with $f_{\rm ede} = 0.09$ for the EDE-- well outside of the 95\% CL posterior distribution.  A similar exploration of the constraints using the EFT-BOSS implementation from Ref.~\cite{Ivanov:2020ril} gives $\chi^2_{\rm EDE} - \chi^2_{\Lambda {\rm CDM}} = -0.016$ with $H_0 =71.94$ km/s/Mpc and $f_{\rm ede} = 0.136$.

\begin{figure*}
\centering
\includegraphics[width=\textwidth]{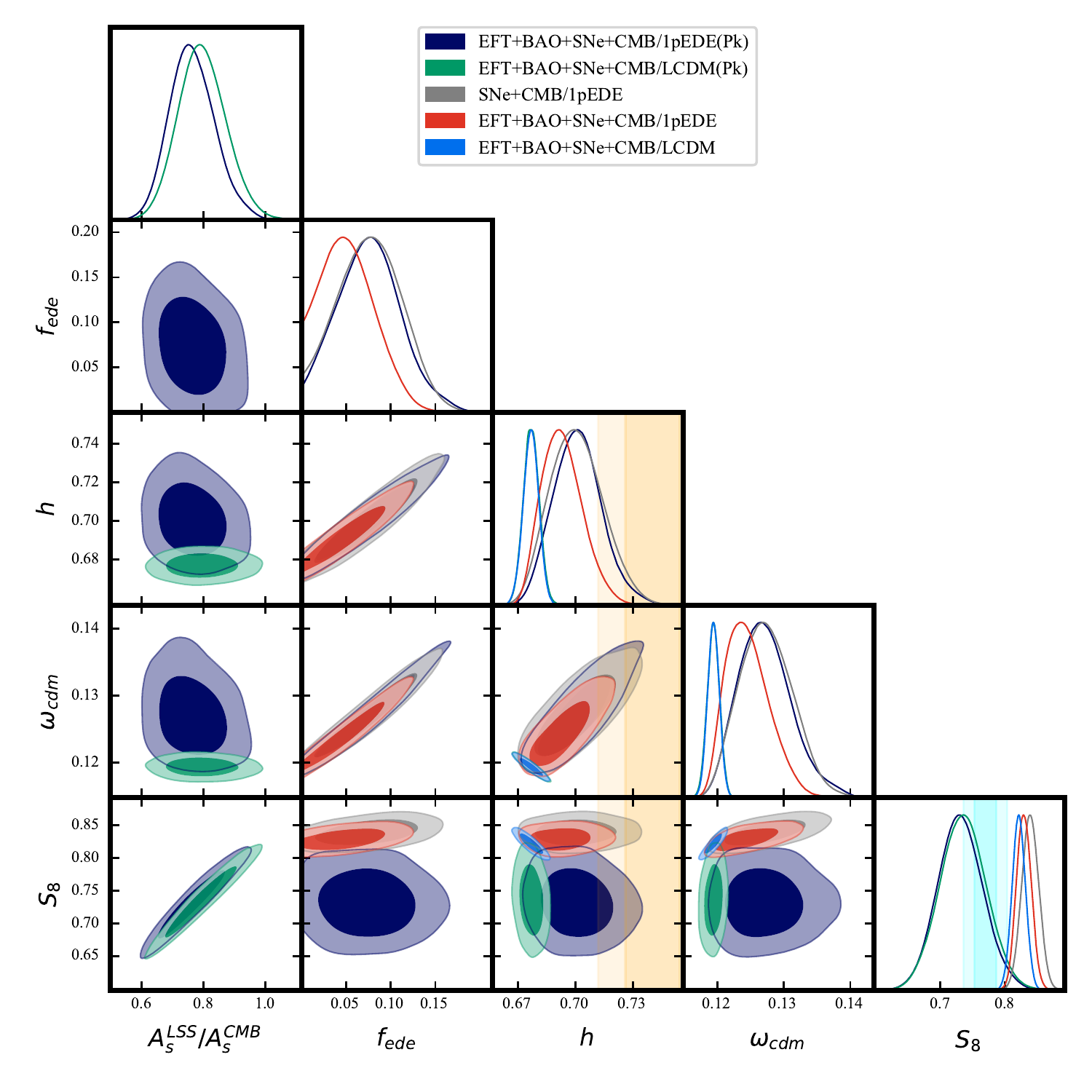}
\caption{68\% and 95\% CL marginalized constraints along with one-dimensional marginalized posteriors. We show results from EFT+BAO+SNe+CMB assuming 1pEDE and 1pEDE(Pk) (red and gray, respectively). We also show results from CMB+SNe assuming 1pEDE in red. The yellow band and blue bands show the 68\% (darker)/95\% (lighter) CL constraints on $h$ from SH0ES \cite{Riess:2019cxk} and on $S_8$ from a combination of weak-lensing measurements (see, e.g., Ref.~\cite{Hill:2020osr}), respectively. Note that for the $(\omega_{\rm cdm},h)$ 2D posteriors the `LCDM' (green) and `LCDM(Pk)' (light blue) lie directly on top of one another.}
\label{fig:split}
\end{figure*}

When claiming to rule out an extension of \LCDM{} it is important to not only rely on the posterior distribution, but also to establish that the extension leads to a degradation in the fit to the data compared to \LCDM. Otherwise, it means that the posterior distributions are driven by the choice of priors. As we have shown here, $f_{\rm ede}$ can take on values much larger than the 95\% CL limits obtained using 3pEDE and still provide as good of a fit to the data. 

\section{Additional constraints on EDE from galaxy clustering}
\label{sec:EDELSS}

Now that we have seen that the inclusion of the EFT+BAO data leads to a decrease in the posterior for the 1pEDE $f_{\rm ede}$ (see Fig.~\ref{fig:progressive_freedom}), it is of interest to establish where this additional constraining power is coming from. 

As a first step, we focus on how EFT+BAO+BBN constrains EDE and compare this with CMB constraints. This comparison is informative for two reasons: first to have a reference to evaluate the constraining power of EFT+BAO; second, to investigate whether there is a degeneracy between parameters that can be potentially broken by the addition of the EFT+BAO likelihood. In all cases, we add SNe to reduce degeneracies.

We show the marginalized constraints from EFT+BAO+BBN and CMB separately for both \LCDM\ and 1pEDE in Fig.~\ref{fig:separate}. As expected, for both data combinations, 1pEDE significantly broadens the marginalized posterior of $h$, as well as shifts it towards higher values. Note that EFT+BAO+BBN+SNe places significantly weaker constraints on $f_{\rm ede}$ than CMB+SNe.

Ref.~\cite{DAmico:2020ods} suggests that the constraining power on EDE from EFT-BOSS comes from a tight constraint on the sound horizon at baryon decoupling, $r_{\rm drag}$, in a joint analysis with \textit{Planck} CMB data. We explore this possibility in Fig.~\ref{fig:separate}, where we show marginalized constraints for the photon-baryon sound horizon at baryon decoupling, $r_{\rm drag}$, in the bottom row. The agreement between the separate EFT-BOSS and \textit{Planck} constraints to the 1pEDE (green and red contours) indicates that in a joint analysis the EFT-BOSS data are not adding further information about $r_{\rm drag}$ and therefore this cannot be the source of the additional constraining power that EFT+BAO+BBN sets on EDE.

On the other hand, Fig.~\ref{fig:separate} shows that for both \LCDM{} and 1pEDE models the EFT+BAO+BBN+SNe data prefer a significantly smaller value for $A_s$ than CMB+SNe,\footnote{We also find a relatively small value of ${\rm ln}(10^{10}A_s)=2.85_{-0.14}^{+0.18}$ (compared to $\ln10^{10}A_{s } = 3.041^{+0.012}_{-0.0088}$ for CMB+SNe) when using the EFT-BOSS implementation from Ref.~\cite{Ivanov:2020ril}.} which may hint at the origin of the strengthened constraints on $f_{\rm ede}$ when including the EFT+BAO+BBN data. 
Indeed, $A_s$ is positively correlated with $f_{\rm ede}$, while EFT+BAO+SNe prefer values of this parameter lower than CMB+SNe, disfavoring high values of $f_{\rm ede}$. 

We can further test this hypothesis by performing a `parameter split' test (see e.g., Ref.~\cite{Bernal:2015zom}). While we keep the standard $A^{\rm CMB}_s$ parameter to fit the CMB power spectra and lensing, we introduce a new parameter, $A_s^{\rm LSS}$, that controls the primordial power spectrum to compute the galaxy power spectrum. The data sets under scrutiny will present a tension between them if $A^{\rm CMB}_s$ is statistically inconsistent with $A_s^{\rm LSS}$. We mark the cases with parameter split adding `(Pk)' to the name of the model.

We show marginalized constraints on $f_{\rm ede}$, $h$ and $S_8$ (computed using `LSS' split parameters) in Fig.~\ref{fig:split}. When we allow for the parameter split, the constraints on $h$ and $f_{\rm ede}$ are similar to those without including EFT+BAO, and $A_s^{\rm LSS}<A_s^{\rm CMB}$ at more than $2\sigma$. This demonstrates that the potential tension between \textit{Planck} and the EFT-BOSS data on $A_s$, present in both EDE and \LCDM{} cosmologies with comparable significance, is the main cause of the gain in constraining power on EDE from the addition of the EFT+BAO likelihood to the analysis.  Interestingly, the $S_8$ marginalized posterior when parameters are split favors lower values than \textit{Planck}, being more consistent with a variety of weak-lensing surveys \cite{Heymans:2013fya,Hildebrandt:2018yau,Abbott:2017wau,Hikage:2018qbn} (shown in the blue bands).

It has also been argued (e.g., Ref.~\cite{Ivanov:2019pdj}) that when EFT-BOSS is combined with \textit{Planck} CMB measurements internal degeneracies are broken leading to an additional constraint on $\omega_{\rm cdm}$. However, it is clear from Figs.~\ref{fig:separate} and \ref{fig:split} that \textit{Planck} constraints on all of the \LCDM{} parameters, except for $A_s$, in the EDE model are in statistical agreement with those from EFT-BOSS.

\section{Conclusions}
\label{sec:cons}

EDE proposes a promising extension to \LCDM\ to resolve the $H_0$ tension and, as any other model, requires strong and robust evidence to be ruled out. While at first sight, LSS observables such as galaxy clustering and weak lensing might seem to provide such evidence, there are two main aspects of current LSS data that challenge  this conclusion. 

First, the EDE parametrization presents an increasingly large prior volume as $f_{\rm ede}\rightarrow 0$, since in this limit the likelihood becomes completely insensitive to variation in the other two EDE parameters. In the posterior distribution, this significantly favors low values of $f_{\rm ede}$, independently of the likelihood. Therefore, given that $f_{\rm ede}$ and the other EDE parameters are uncorrelated in the region of the parameter space of interest, by keeping $f_{\rm ede}$ as the only free parameter (our 1pEDE model), we produce posteriors that more directly samples the likelihood. We have shown that in this case, the constraints on $f_{\rm ede}$  weaken significantly. 

Second, we have also identified the origin of the additional constraining power provided by the EFT+BAO likelihood to a joint analysis. It is the small tension between \textit{Planck} and EFT+BAO inferred values of $A_s$ (the latter favoring a lower value), together with the positive correlation between $f_{\rm ede}$ and $A_s$, which places stronger constraints on $f_{\rm ede}$. We have demonstrated this by performing a `parameter split' test, allowing $A_s$ for the LSS data to vary independently of $A_s$ for the CMB power spectra and lensing. Given the tension between EFT+BAO and \textit{Planck} data, even when analyzed using the \LCDM{} model (see Fig.~\ref{fig:split}), one should be cautious when interpreting constraints to models beyond LCDM{} using a joint analysis. 

Our results point out that, given the discussion above, there is not enough evidence to rule EDE models out. First, as shown in Fig.~\ref{fig:separate}, EDE, with $f_{\rm ede}>0$, can provide a good fit to \textit{Planck} and BOSS galaxy power spectra, separately. Second, the EDE cosmology with non-zero $f_{\rm ede}$ can provide as good of a joint fit to \textit{Planck} and BOSS galaxy power spectra as \LCDM. For example, we find that when fixing $H_0= 71$ km/s/Mpc in the 3pEDE model, $\chi_{\rm EDE}^2-\chi_{\Lambda{\rm CDM}}^2 = 0.12$, with $f_{\rm ede} = 0.09$.

Even though our analysis has been done exclusively with a particular EDE model, our conclusions apply more broadly to any extension to the \LCDM{} model whose main impact is to increase the pre-recombination expansion rate due to a material with significant internal pressure support. The parameter controlling the size of such an increase will have a positive correlation with $A_s$ (in order to fit CMB measurements) and will therefore be impacted by the EFT-BOSS data in a similar way (e.g., Refs.~\cite{Niedermann:2020qbw,Lin:2020jcb}). 

Although current observations do not provide strong evidence for or against EDE models, forthcoming CMB and LSS experiments are expected to be precise enough to resolve this ambiguity. As discussed in Refs.~\cite{Smith:2019ihp,Ivanov:2020ril,Klypin:2020tud}, CMB-S4, Euclid/DESI-like spectroscopic galaxy surveys, and James Webb Space Telescope observations of galaxy abundances and clustering, should be able to definitively probe these predictions, and future spectral distortion measurements may test the high values of $n_s$ required by EDE~\cite{Lucca:2020fgp}. Moreover, EDE might entail unique predictions regarding the production of chiral gravitational waves, scalar, and possibly vector perturbations \cite{Berghaus:2019cls, Gonzalez:2020fdy, Weiner:2020sxn}, or have impact in the neutrino mass~\cite{Sakstein:2019fmf} or cosmological light scalar fields~\cite{Griest:2002cu}. Our work shows that, contrary to the claims that the EDE model has been `ruled out', the analysis of the various predictions of EDE with future data is still warranted.

\acknowledgments
We thank Tanvi Karwal for helpful discussions, Guido D'Amico and Pierre Zhang for help with running {\sf PyBird}. This work used the Strelka Computing Cluster, which is run by Swarthmore College. TLS is supported by NSF Grant No.~2009377, NASA Grant No.~80NSSC18K0728, and the Research Corporation.  This work was supported at Johns Hopkins by NSF Grant No.\ 1818899 and the Simons Foundation.  JLB is supported by the Allan C.\ and Dorothy H.\ Davis Fellowship.   

\bibliography{bibliography}

\end{document}